\documentclass[amssymb,amsmath,aps,showpacs,floatfix,nofootinbib,showpacs,12pt]{revtex4}
\usepackage{epsfig}
\usepackage{color}
\usepackage{graphicx}

\def\beq{\begin{equation}}
\def\eeq{\end{equation}}

\def\be{\begin{equation}}
\def\ee{\end{equation}}
\def\bea{\begin{eqnarray}}
\def\eea{\end{eqnarray}}

\newcommand{\gsim}{\lower.7ex\hbox{$\;\stackrel{\textstyle>}{\sim}\;$}}
\newcommand{\lsim}{\lower.7ex\hbox{$\;\stackrel{\textstyle<}{\sim}\;$}}

\begin{document}
\title{A Light Sterile Neutrino from Friedberg-Lee Symmetry}

\author{ Xiao-Gang He$^{1,2,3,4}$, Wei Liao$^{5}$
\footnote{Correspondence to: liaow@ecust.edu.cn}}
\affiliation{ 
$^1$Shanghai Key Laboratory of Particle Physics and Cosmology, Shanghai\\
${}^{2}$INPAC and Department of Physics,
Shanghai Jiao Tong University, Shanghai}
\affiliation{${}^{3}$Physics Division, National Center for
Theoretical Sciences, Department of Physics, National Tsing Hua
University, Hsinchu} 
\affiliation{${}^{4}$CTS, CASTS and
Department of Physics, National Taiwan University, Taipei}
\affiliation{${}^5$Institute of Modern Physics, School of Science,
East China University of Science and Technology, Shanghai
}

\begin{abstract}
Light sterile neutrinos of mass about an eV with mixing 
$\tilde U_{ls}$ of a few percent to active neutrinos
may solve some anomalies shown in experimental data related 
to neutrino oscillation. How to have light sterile neutrinos is 
one of the theoretical problems which have attracted 
a lot of attentions. In this article we show that such an eV scale light 
sterile neutrino candidate can be obtained in a seesaw model in which the 
right-handed neutrinos satisfy a softly-broken Friedberg-Lee (FL) symmetry.
In this model a right-handed neutrino is guaranteed by the FL symmetry to 
be light comparing with other two heavy right-handed neutrinos.
It can be of eV scale when the FL symmetry is softly broken and
can play the role of eV scale sterile neutrino needed for explaining the 
anomalies of experimental data. This model predicts that one 
of the active neutrino is massless. 
We find that this model prefers
inverted hierarchy mass pattern of active neutrinos than normal hierarchy.
An interesting consequence of this model is that realizing
relatively large $|{\tilde U}_{es}|$ and relatively small $|{\tilde U}_{\mu s}|$
in this model naturally leads to a relatively small $|{\tilde U}_{\tau s}|$.
This interesting prediction can be tested in future atmospheric or solar
neutrino experiments.

\end{abstract}


\maketitle

 \noindent {\bf Introduction}

Experiments have confirmed the existence of three active neutrinos, 
$\nu_e$, $\nu_\mu$ and $\nu_\tau$ participating the usual weak interactions and mixing with each other,
beyond reasonable doubt. Many of the experimental data on neutrino 
oscillation can be explained by mixing among these neutrinos and 
the mixing angles have been determined~\cite{review}. 
Various data, such as the 
invisible decay width of the Z boson, have excluded the existence of 
a fourth light active neutrino. However, additional light sterile neutrinos 
which do not participate the usual weak interactions, but may mix with the 
active ones, have not been excluded. In fact there are several experimental 
indications showing that sterile neutrinos may help to solve some problems 
show in experimental data. These problems include anomalies show in data from the LSND appearance 
experiment~\cite{LSND}, the MiniBooNE neutrino and anti-neutrino 
appearance experiments~\cite{MiniBooNE}, 
the reactor neutrino flux anomaly data~\cite{reactorAnomaly}, and 
the data from the deficit of neutrino spectrum in Gallium radioactive source
experiment~\cite{galliumAnomaly}. 
If the sterile neutrinos have masses of order an eV, some or all these problems 
can be resolved~\cite{sterilefit1,sterilefit2,sterilefit3,sterilefit4}. 
In view of its possible solution to these problems, 
an eV scale sterile neutrino, although not favored by some other experiments
~\cite{KARMEN,NOMAD} and the tension between appearance and 
disappearance experiments~\cite{sterilefit3},
has raised great interests of particle physicists~\cite{sterilenuref}
with several experiments proposed to test 
the existence of sterile neutrinos~\cite{sterilenuproposal,icecube}.
How to have light 
sterile neutrinos is one of theoretical problems which have attracted 
a lot of attentions. In this article we discuss such an eV scale light 
sterile neutrino candidate in seesaw model~\cite{seesaw1} in which the 
right-handed neutrinos satisfy the Friedberg-Lee (FL) symmetry. We show that 
in this model, one can naturally have a light neutrino. With soft-breaking 
of the FL symmetry, we find parameter spaces which can explain
preferred sterile neutrino mass and mixing.

\vspace{0.3cm}
\noindent {\bf Friedberg-Lee symmetry and neutrino mass pattern}

We will work with type I seesaw model with 3 active neutrinos which 
belong to electroweak doublet $L_{Li} = (\nu_{Li}, e_{Li})^T$ and 
3 right-handed neutrinos $\nu_{Ri}$ which transform as singlets
under the SM gauge group. 
The Lagrangian responsible to neutrino masses is
 \begin{eqnarray}
{\cal L} =  - {1\over 2}
\bar \nu_R M \nu^c_R  - \bar L_L Y H \nu_R + H.C. \;,\label{final-form}
\end{eqnarray}
where $H = (H^0, H^-)^T$ is the Higgs doublet. $\nu^c_{R}$ is the charge conjugate of $\nu_R$. $M$ and $Y$ are $3\times 3$ matrices. $M$ is the Majorana mass
 matrix of $\nu_R$ and is symmetric. 

After the electroweak symmetry breaking, that is, the Higgs develops
a non-zero vacuum expectation value $\langle H\rangle = (v,0)^T$,
the neutrino mass matrix
 in the basis $(\nu_L, \nu_R^c)^T$ is given by
 \begin{eqnarray}
 \left (\begin{array}{cc} 0&Y^*v \\Y^\dagger v &M \end{array} \right )\;.
 \end{eqnarray}

The usual seesaw model assumes that $M$ is rank 3 and the eigenvalues are 
much larger than the electroweak scale to obtain light neutrino 
masses of order 0.1 eV or smaller. Without additional assumptions, 
there is no light right-handed sterile neutrinos. A possible scenario 
of having a light sterile neutrino is to impose a symmetry to the model 
which leads to a massless right-handed neutrino (or neutrinos) 
to be identified as the light sterile neutrino and 
to induce a finite small mass by softly breaking this symmetry.
In Ref.~\cite{HeLiao1} it was shown that
an exact global Friedberg-Lee(FL) symmetry in the right-handed neutrino sector
implies that one right-handed neutrino is massless and decoupled from
other neutrinos. In Ref.~\cite{HeLiao2} it was argued that an 
approximate FL symmetry in the right-handed neutrino sector implies 
that one right-handed neutrino can be very light comparing with
other right-handed neutrinos. Therefore a seesaw model accessed with 
the FL symmetry may provide a natural way to obtain a light sterile neutrino. 
In this work, we carry out a detailed analysis to show how to realize an
eV scale sterile neutrino in this scenario 
and discuss possible interesting consequences. 

To start with, let us briefly 
review how a FL symmetry can lead to a massless sterile neutrino.
A theory is said to have a FL symmetry when the Lagrangian of this theory
is invariant under a transformation on a fermionic field of the form
 $q \to q + \epsilon$~\cite{Friedberg:2006it, Friedberg:2007ba,others}, 
where $\epsilon$ is a space-time independent element of the Grassmann
algebra, anti-commuting with the fermionic field operators $q$. 
Imposition of a FL symmetry for the SM particles which actively 
participate in electroweak interactions may be too restrictive 
for the theory to survive known experimental constraints. 
For right-handed neutrinos they may allow such a possibility. 
One can have the FL symmetry along a particular direction in right-handed 
flavor space, $q = \xi_1 \nu_{R1} + \xi_2 \nu_{R2}+\xi_3 \nu_{R3}$ and 
require the theory to be invariant under $q\to q+\epsilon$ transformation. 
By making an appropriate transformation in flavour space, 
relabelling $\nu_{R1}$ to be $q$, and the other orthogonal 
states to be $\nu_{R2,3}$, the invariance of the Lagrangian in 
Eq.(\ref{final-form}) under a global FL transformation 
becomes the invariance of the Lagrangian under
 \begin{eqnarray}
 \nu_{R_1}\to \nu_{R_1}+\epsilon\;. \label{FL}
 \end{eqnarray}

It is easy to check that the kinetic term
 ${\cal L}_k = \bar \nu_R \gamma_\mu (i \partial^\mu \nu_R)$ is
 invariant under a transformation defined in Eq.(\ref{FL}) up
 to a total derivative.
The invariance of the Yukawa coupling term under 
the transformation Eq.(\ref{FL}) gives
 \begin{eqnarray}
 Y= \begin{pmatrix} 0 & {\tilde Y}_{e 2} & {\tilde Y}_{e 3} \cr
 0 & {\tilde Y}_{\mu 2} & {\tilde Y}_{\mu 3} \cr
 0 & {\tilde Y}_{\tau 2} & {\tilde Y}_{\tau 3} \cr
 \end{pmatrix}. \label{Y-FL}
 \end{eqnarray}
 
The invariance of the Majorana mass term under 
the transformation Eq.(\ref{FL}) gives
 \begin{eqnarray}
 M = \begin{pmatrix} 
0 & 0 & 0 \cr 
0 & {\tilde M}_{22} &  {\tilde M}_{23} \cr
0 & {\tilde M}_{23} & {\tilde M}_{33} \cr  
\end{pmatrix}\;.
 \label{M-FL}
 \end{eqnarray}
 
Note that the Majorana mass matrix is forced to be a rank two matrix. 
$\nu_{R1}$ is massless. The non-zero eigenvalues of $M$ are
heavy to facilitate the seesaw mechanism.

One can integrate out two heavy neutrinos $\nu_{R2,R3}$ and
get a mass matrix for $(\nu_e,\nu_\mu,\nu_\tau, \nu^c_{R1})$:
\bea
m_\nu= \begin{pmatrix} 
{\tilde m}_\nu & 0_{3\times 1} \cr
0_{1\times 3} & 0 \cr
\end{pmatrix}, \label{numass0a}
\eea
where ${\tilde m}_\nu$ is a $3\times 3$ matrix:
\bea
{\tilde m}_\nu=- {\tilde Y}^* {\tilde M}^{-1} {\tilde Y}^\dagger v^2.
\label{numass0}
\eea
${\tilde Y}$ and ${\tilde M}$ are
\bea
{\tilde Y} &&= \begin{pmatrix} {\tilde Y}_{e 2} & {\tilde Y}_{e 3} \cr
  {\tilde Y}_{\mu 2} & {\tilde Y}_{\mu 3} \cr
 {\tilde Y}_{\tau 2} & {\tilde Y}_{\tau 3} \cr
 \end{pmatrix}, \label{Y0} \\
 {\tilde M} && = \begin{pmatrix} 
{\tilde M}_{22} &  {\tilde M}_{23} \cr
{\tilde M}_{23} & {\tilde M}_{33} \cr  
\end{pmatrix}. \label{M0}
\eea
${\tilde m}_\nu$ is a rank two matrix which gives
two non-zero neutrino masses. One combination of
active neutrinos is massless in this model.
In this scenario, masses and mixings of low energy neutrinos are given
by seesaw mechanism with two heavy right-handed neutrinos,
a scenario called the minimal seesaw~\cite{Frampton:2002qc}.
We see that an exact FL symmetry in right-handed neutrino sector
reduces the usual seesaw model to the minimal seesaw model~\cite{HeLiao1}.

It is easy to see in Eq. (\ref{numass0a}) that this scenario gives a 
massless right-handed neutrino which decouples from all other neutrinos.
It can not provide a low energy sterile neutrino which mixes
with active neutrinos. Deviation from or breaking of FL symmetry
introduced in Eq. (\ref{FL}) is needed to accommodate an eV scale sterile
neutrino which mixes with active light neutrinos to solve some of the 
problems mentioned earlier.

\vspace{0.3cm}
 \noindent {\bf Low energy sterile neutrino with soft-breaking FL}

In this section we discuss how soft FL symmetry breaking can help to make a realistic model.
Soft breaking of FL symmetry can only occur in the Majorana mass sector $M$. 
With soft-breaking terms of FL symmetry, the Majorana mass 
matrix can be written as
\bea
 M = \begin{pmatrix} 
x_{11} & x_{12} & x_{13} \cr 
x_{12} & {\tilde M}_{22} &  {\tilde M}_{23} \cr
x_{13} & {\tilde M}_{23} & {\tilde M}_{33} \cr  
\end{pmatrix}.
 \label{M-approxFL}
\eea

Non-zero values of $x_{1i}$ softly break the FL symmetry. 
Since these terms break the FL symmetry, they are naturally much 
smaller than the eigenvalues of ${\tilde M}_{ij}$ according to 't Hooft
naturalness condition. The actual values of $x_{1i}$ are not known. 
We will take them as free parameters to be determined or 
constrained by experimental data. 

After integrating out two heavy neutrinos
 we get a mass matrix for 
 $(\nu_e,\nu_\mu,\nu_\tau, \nu^c_{R1})$:
 \bea
 m_\nu= \begin{pmatrix} 
 {\tilde m}_\nu & -{\tilde Y}^* v {\tilde M}^{-1} X^T \cr
 -X {\tilde M}^{-1} {\tilde Y}^\dagger  v  & 
x_1 \cr
 \end{pmatrix}, \label{numass2}
 \eea
where
\bea
x_1=x_{11}- X {\tilde M}^{-1} X^T,
\eea
and $X=(x_{12},x_{13})$.
${\tilde m}_\nu$ and ${\tilde M}$ have been given in
Eqs. (\ref{numass0}) and (\ref{M0}).

In the limit that $x_{12, 13}$ are zero, $x_{11}$ is the 
sterile neutrino mass $m_{\nu_s}$ which we assume to be of order eV. 
In this case, only active neutrinos mix with each other and
the light neutrino mass matrix in Eq.(\ref{numass2}) is diagonalized by
\bea
{\tilde U}=
\begin{pmatrix} U & 0 \cr 0 & 1 \cr \end{pmatrix}\;,
 \label{mixing0}
\eea
where $U$ is the usual PMNS mixing for active neutrinos defined 
by $\tilde m_\nu = U^*\tilde m^\prime_\nu U^\dagger$. 
Here ${\tilde m}'_\nu=\textrm{diag}\{0,m_2,m_3\}$ for normal hierarchy(NH)
of light active neutrino masses and 
${\tilde m}'_\nu=\textrm{diag}\{m_1,m_2,0\}$ for inverted hierarchy(IH).
We will work with the convention that the Majorana phases are kept in 
the mass eigenvalues, and therefore, $U$ does not contain any Majorana phases.

In this case a general expression
for ${\tilde Y}$ which can produce the desired NH neutrino mass
pattern can be written as follows~\cite{HeLiao2}
\bea
 {\tilde Y} v= i U ({\tilde m}^{'*}_\nu)^{1/2}
 \begin{pmatrix} 0 & 0 \cr
                 \cos \theta & \sin\theta \cr
                 -\sin \theta & \cos\theta
 \end{pmatrix}  ({\tilde M}^{'*})^{1/2}, \label{NHmatrix1}
 \end{eqnarray}
where $\theta$ is a complex number, 
and ${\tilde M}'=\textrm{diag} \{M_2,M_3 \}$ is a diagonalized mass matrix 
for heavy neutrinos. Without loss of generality we can diagonalize 
${\tilde M}$ and make discussion in this base. Using (\ref{numass0}) 
and ${\tilde M}'$ one can easily check that (\ref{NHmatrix1}) 
reproduces the NH neutrino mass matrix.
 
For IH, ${\tilde m}'_\nu=\textrm{diag}\{m_1,m_2,0\}$ and we have
~\cite{HeLiao2}
\bea
 {\tilde Y} v= i U ({\tilde m}^{'*}_\nu)^{1/2}
 \begin{pmatrix}
                 \cos \theta & \sin\theta \cr
                 -\sin \theta & \cos\theta \cr
                 0&0
 \end{pmatrix}  ({\tilde M}^{'*})^{1/2} . \label{IHmatrix1}
 \end{eqnarray}
 
When $x_{12, 23}$ become non-zero, mixing between active and 
the light sterile neutrino will happen. One can approximate, in general, 
the mixing matrix for small active-sterile neutrino mixing
as follows
 \bea
{\tilde U}=
\begin{pmatrix} 1 & R \cr -R^\dagger & 1\cr \end{pmatrix}
\begin{pmatrix} U & 0 \cr 0 & 1 \cr \end{pmatrix}
, \label{mixing1}
\eea
where $R$ is a $3\times 1$ matrix representing the
mixing of active neutrinos and sterile neutrino. The above expression is valid
as long as $R_{ls}$($l=e,\mu,\tau$), the element of $R$, 
satisfies $|R_{ls}|^2 \ll 1$. Diagonalizing the fourth
row and the fourth column in Eq. (\ref{numass2}) using the
first matrix in Eq. (\ref{mixing1}) we find that 
$R$ is solved as
\bea
R^*\approx -\frac{1}{x_1} {\tilde Y}^* v {\tilde M}^{-1} X^T  ,
\label{mixingR}
\eea
and
the neutrino mass matrix becomes
\bea
\begin{pmatrix}
{\hat m}_\nu= {\tilde m}_\nu-R^* x_1 R^\dagger & 0 \cr
0 & x_1 \cr 
\end{pmatrix}, \label{numass3}
\eea
where order $R^\dagger R$ correction to $x_1$ has been neglected.
${\hat m}_\nu$ in Eq. (\ref{numass3}) is further diagonalized
using $U$ in the second matrix in Eq. (\ref{mixing1}) with
$U^T {\hat m}_\nu U=\textrm{diag}\{m_1,m_2,m_3\}$.
The order $R^\dagger R$ correction to the sterile neutrino mass
from active and sterile neutrino mixing can be neglected
for $|R_{ls}|^2 \ll 1$ and we have
\bea
m_s\approx x_1= x_{11}-X {\tilde M}^{-1} X^T. \label{sterilenumass}
\eea

If $x_{12,13}$ are of the order as $x_{11}$, {\it i.e.} of order eV or
 tens eV,
they cannot provide any explanation for the anomalies mentioned 
earlier since the mixing of the sterile neutrino with the active ones 
will be very small, as can be seen in Eq. (\ref{mixingR}). 
To have the mixing to be of order of interests, say, 
about 0.1, $x_{12,13}$ should satisfy 
${\tilde Y} v \tilde M^{* -1} X^\dagger/x_1 \approx 0.1$. 
With $x_{1}$ of order in the eV range, elements in $X$ should be 
an order of magnitude larger than elements in $\tilde Y v$. 
In this case the contribution to sterile neutrino mass from $x_{12,13}$ 
may not be neglected. In order that there is no fine-tuning
of two terms in Eq. (\ref{sterilenumass}) greater than 
$1\%$ level, we get that $x_{12,13} \lsim 10 \sqrt{\textrm{eV}~ M_{2,3}}$.
For this range of the magnitude of $X$, it's sufficient to get
a mixing of active-sterile neutrinos of $\gsim 0.1$.
The hierarchy for various quantities are therefore: 
$\tilde Y v < X \ll \tilde M$.

This estimate of the order of magnitude of $x_{12,13}$ may also come 
from considerations of how large the soft-breaking terms should be. 
Since the soft-breaking terms are all related to right-handed neutrino 
mass matrix, a reasonable criteria for the size of the soft-breaking 
terms is that the smallest eigenvalue should be much smaller than 
two large eigen-masses already existed when the soft-breaking terms are
absent. The lightest eigenvalue, phenomenologically, should be 
$m_s$, the sterile neutrino mass, which is about an eV or so 
to be of interests. We therefore take that as a requirement. 
In the case that this requirement is satisfied
the lightest eigenvalue of $M$ can be computed as
\bea
m_s=det(M)/det({\tilde M})= x_{11}-X {\tilde M}^{-1} X^T, \label{sterilenumass1}
\eea
which is consistent with Eq. (\ref{sterilenumass}).
One can see in Eq. (\ref{sterilenumass1}) that this requirement only 
limit $m_s$ to be of order eV, 
but still allow $x_{12,13}$ to be larger 
since its contribution to the mass eigenvalues are of order 
$X {\tilde M}^{-1} X^T$. On the other hand, the order of 
magnitude of $x_{12,13}$ should 
be restricted to much smaller than large non-zero masses in ${\tilde M}$.
When these conditions are satisfied, the soft-breaking scale of FL symmetry 
can be considered to be natural although
$x_{12,13}$ can be orders of magnitude different from $x_{11}$. 
We will work with the approximation 
conditions described above and turn to discuss realizing active
neutrino mixing in this scenario.

Using Eq. (\ref{mixingR}), one can see in Eq. (\ref{numass3}) that 
the mass matrix giving rise to the
PMNS matrix is no longer Eq. (\ref{numass0}). It is
\bea
{\hat m}_\nu &&= {\tilde m}_\nu-R^* x_1 R^\dagger
= - {\tilde Y}^* v {\tilde M}^{-1/2} S {\tilde M}^{-1/2} {\tilde Y}^\dagger v,
\label{numass4}
\eea
where
\bea
S =1+\frac{1}{x_1} {\tilde M}^{-1/2} X^T X {\tilde M}^{-1/2}.
\label{numass4a}
\eea
Since $x_1$ is of order eV and $|R_{ls}| \sim 0.1$, the correction
to ${\tilde m}_\nu$ due to mixing $R$ can not be neglected.
Although the mass matrix Eq. (\ref{numass4}) is more complicated 
than Eq. (\ref{numass0}),
it is still rank two and has one zero eigenvalue. This can
be clearly seen in Eq. (\ref{numass4}) by noting that ${\tilde Y}$
is rank two. One can also see this point in Eq. (\ref{numass2})
by noting that the first to third rows are proportional to ${\tilde Y}^*$
which is rank two and the total matrix is rank three.

Similar to Eq. (\ref{NHmatrix1}) we can obtain an expression of
${\tilde Y}$ for NH
\bea
 {\tilde Y} v= i U ({\tilde m}^{'*}_\nu)^{1/2}
 \begin{pmatrix} 0 & 0 \cr
                 \cos \theta & \sin\theta \cr
                 -\sin \theta & \cos\theta
 \end{pmatrix} ({\hat S}^*)^{-1/2} \Lambda^\dagger
({\tilde M}^{'*})^{1/2}, \label{NHmatrix2}
 \end{eqnarray}
where
${\hat S}$ is a diagonalized matrix 
and $\Lambda$ is a unitary matrix which diagonalizes matrix $S$:
\bea
\Lambda^T S \Lambda ={\hat S}. \label{S}
\eea
The mixing of active neutrinos with sterile neutrino $R$ is expressed as
\bea
R^*\approx \frac{i}{x_1} U^* ({\tilde m}^{'}_\nu)^{1/2}
\begin{pmatrix} 0 & 0 \cr
                 (\cos \theta)^* & (\sin\theta)^* \cr
                 -(\sin \theta)^* & (\cos\theta)^* \end{pmatrix}
{\hat S}^{-1/2} \Lambda^T ({\tilde M}^{'})^{-1/2} X^T,
\label{mixingR1}
\eea
For IH an expression similar to Eq. (\ref{mixingR1}) can be obtained:
\bea
R^*\approx \frac{i}{x_1} U^* ({\tilde m}^{'}_\nu)^{1/2}
\begin{pmatrix} 
                 (\cos \theta)^* & (\sin\theta)^* \cr
                 -(\sin \theta)^* & (\cos\theta)^* \cr
   0 & 0 \cr      \end{pmatrix}
{\hat S}^{-1/2} \Lambda^T ({\tilde M}^{'})^{-1/2} X^T.
\label{mixingR2}
\eea
Introducing 
${\hat X}^T=\frac{1}{\sqrt{x_1}} {\tilde M}^{-1/2} X^T$,
$S$ can be re-expressed as $S=1+{\hat X}^T {\hat X}$.
A simple scenario is when ${\hat X}$ is real. In this case,
a $\Lambda$ diagonalizing $S$ is
\bea
\Lambda={\hat X}^T_1 \begin{pmatrix} 1 & 0 \cr \end{pmatrix}
+{\hat X}_2^T \begin{pmatrix} 0 & 1 \cr \end{pmatrix},
\eea
where ${\hat X}_{1,2}$ are two normalized $1\times 2$ real matrices and
they satisfy: ${\hat X}_i {\hat X}^T_j=\delta_{ij} (i,j=1,2)$ and
${\hat X}_2 {\hat X}^T =0$.
It is easy to check that $\Lambda^T S \Lambda 
=\textrm{diag} \{1+{\hat X} {\hat X}^T, 1\}$,
\bea
R^*\approx \frac{i}{\sqrt{x_1}} U^* ({\tilde m}^{'}_\nu)^{1/2}
\begin{pmatrix} 0 & 0 \cr
                 (\cos \theta)^* & (\sin\theta)^* \cr
                 -(\sin \theta)^* & (\cos\theta)^* \end{pmatrix}
\begin{pmatrix} \sqrt{{\hat X}{\hat X}^T/(1+{\hat X}{\hat X}^T)} \cr
0 \cr \end{pmatrix}
\label{Rmatrix1}
\eea
for NH, and
\bea
R^*\approx \frac{i}{\sqrt{x_1}} U^* ({\tilde m}^{'}_\nu)^{1/2}
\begin{pmatrix}
                 (\cos \theta)^* & (\sin\theta)^* \cr
                 -(\sin \theta)^* & (\cos\theta)^* \cr
  0 & 0 \cr        \end{pmatrix}
\begin{pmatrix} \sqrt{{\hat X}{\hat X}^T/(1+{\hat X}{\hat X}^T)} \cr
0 \cr \end{pmatrix}
\label{Rmatrix2}
\eea
for IH.

\vspace{0.3cm}
 \noindent {\bf Numerical Analysis}

In this section, we will study some implications of the 
light neutrino mass matrix discussed in the previous section 
resulting from soft-breaking FL symmetry with sterile neutrino
mass of order eV and sterile-active neutrino mixing of order 10\%. 
For an illustration of our scenarios we will try to obtain the best fit
of the sterile neutrino mass and mixing~\cite{sterilefit1}:
\bea
m_s=1.27 ~\textrm{eV},~~|{\tilde U}_{es}|^2=0.035,
~~|{\tilde U}_{\mu s}|^2=0.0086, \label{parameters}
\eea
where $|{\tilde U}_{\mu s}|$ is considerably smaller than $|{\tilde U}_{e s}|$
given by the null result of short-baseline $\nu_\mu$ disappearance experiment.
${\tilde U}_{\tau s}$ is constrained by atmospheric and solar neutrino data
~\cite{sterilefit3}:
\bea
|{\tilde U}_{\tau s}|^2 < 0.2, ~~\textrm{2 $\sigma$}. \label{parametersb}
\eea

In our numerical analysis, we use the neutrino mass squared differences 
and mixing of active neutrinos as the following~\cite{globalfit2}
\begin{eqnarray}
&& \Delta m^2_{21}= 7.62\times 10^{-5} ~\textrm{eV}^2,~~
|\Delta m^2_{31}|=2.50\times 10^{-3} ~\textrm{eV}^2, \label{parameters1} \\
&& \sin^2\theta_{12}=0.32,~~\sin^2\theta_{23}=0.60,
~~\sin^2\theta_{13}=0.025. \label{parameters2}
\end{eqnarray}
Since mass squared differences and mixing angles for NH and IH
are almost the same~\cite{globalfit2} we neglect the differences
for these two mass patterns and use Eqs. (\ref{parameters1}) 
and (\ref{parameters2}) for both cases.

Since one of the active neutrino mass is zero in both NH and IH cases, 
the neutrino mass are all known. We find solutions for our scenarios:
\begin{eqnarray}
&&\mbox{NH}: m_1=0, ~~|m_2|=\sqrt{\Delta m^2_{21}}\approx 0.873\times 10^{-2} ~\textrm{eV},~~|m_3|\approx \sqrt{|\Delta m^2_{31}|}\approx 0.05 ~\textrm{eV}, 
\label{parameters3}\\
&&\mbox{IH}: |m_1|=\sqrt{|\Delta m^2_{31}|}\approx 0.05 ~\textrm{eV},
~~|m_2|\approx\sqrt{|\Delta m^2_{31}|-\Delta m^2_{21}}\approx 0.05 ~\textrm{eV},
~~m_3=0.  \label{parameters4}
\end{eqnarray}
The mixing matrix $U$ is expressed using $\theta_{ij}$ as follows
\bea
U=\begin{pmatrix}c_{12} c_{13} &  s_{12} c_{13} & s_{13}e^{-i\delta} \cr
-s_{12}c_{23}-c_{12}s_{23}s_{13}e^{i\delta} & 
c_{12}c_{23}-s_{12}s_{23}s_{13}e^{i\delta} & s_{23} c_{13} \cr
s_{12}s_{23}-c_{12}c_{23}s_{13}e^{i\delta} & 
-c_{12}s_{23}-s_{12}c_{23}s_{13}e^{i\delta} & c_{23} c_{13} \cr \end{pmatrix},
\label{Umatrix}
\eea
where $s_{ij}=\sin\theta_{ij}$, $c_{ij}=\cos\theta_{ij}$ and
$\delta$ is a CP violating phase.

For NH we can read in Eq. (\ref{Rmatrix1}) 
\bea
R\approx \frac{-i}{\sqrt{m_s}}\sqrt{\frac{x_{11}-x_1}{x_{11}}}
\begin{pmatrix}
 U_{e2} (m_2^*)^{1/2} \cos\theta -U_{e3} (m_3^*)^{1/2} \sin\theta \cr
 U_{\mu 2} (m_2^*)^{1/2} \cos\theta - U_{\mu 3} (m_3^*)^{1/2} \sin\theta \cr
 U_{\tau 2} (m_2^*)^{1/2} \cos\theta -U_{\tau 3} (m_3^*)^{1/2} \sin\theta \cr
 \end{pmatrix},
\label{Rmatrix3}
\eea
where ${\hat X}{\hat X}^T/(1+{\hat X}{\hat X}^T)=(x_{11}-x_1)/x_{11}$
has been used. In this scenario it's difficult to have larger
$|{\tilde U}_{es}|^2$($|R_{e s}|^2$)
and smaller $|{\tilde U}_{\mu s}|^2$($|R_{\mu s}|^2$) 
as shown in Eq. (\ref{parameters}). One can see this by noting that
$|m_3/m_s|\approx 0.0394$ and $|U_{e3}|<|U_{\mu 3}|$ contrary to
the associated hierarchy of $|R_{e s}|$ and $|R_{\mu s}|$. So suppression of
contributions of $\sqrt{m_3^*/m_s}$ in $|R_{e s}|$ and $|R_{\mu s}|$ is
needed. This can be achieved by taking $|\sin\theta|<1$. Unfortunately
we can find that $|m_2/m_s|\approx 0.00685$ and for $|R_{es}|^2$ to
reach $0.035$ we need $|cos\theta|^2 \gg 1$. These two requirements
on $\cos\theta$ and $\sin\theta$ are hard to reconcile even allowing 
complex $\theta$.

For IH we can read in Eq. (\ref{Rmatrix2}) 
\bea
R\approx \frac{-i}{\sqrt{m_s}}\sqrt{\frac{x_{11}-x_1}{x_{11}}}
\begin{pmatrix}
 U_{e1} (m_1^*)^{1/2} \cos\theta -U_{e2} (m_2^*)^{1/2} \sin\theta \cr
 U_{\mu 1} (m_1^*)^{1/2} \cos\theta - U_{\mu 2} (m_2^*)^{1/2} \sin\theta \cr
 U_{\tau 1} (m_1^*)^{1/2} \cos\theta -U_{\tau 2} (m_2^*)^{1/2} \sin\theta \cr
 \end{pmatrix},
\label{Rmatrix4}
\eea
Since $|m_1/m_s|\approx |m_2/m_s|\approx 0.0394$ for IH,
their contributions to $R_{ls}$ are equally important.
Suppression of $R_{\mu s}$ can be achieved by making
two terms proportional to $U_{\mu 1}$ and $U_{\mu 2}$
in $R_{\mu s}$ are of opposite signs and cancel with each other
while two terms proportional to $U_{e 1}$ and $U_{e 2}$ 
in $R_{es}$ are of the same sign. This is possible because in our convention
$U_{e1} U_{e2}=\cos^2\theta_{13}\sin\theta_{12}\cos\theta_{12}$ and 
$U_{\mu 1}U_{\mu 2}\approx -\cos^2\theta_{23}\sin\theta_{12}\cos\theta_{12}$
which is exactly the case we want. An example to realize 
Eq. (\ref{parameters}) is $\delta=\pi$,
$(m_2^*/m_1^*)^{1/2}\approx -i$, $\cos\theta=\sqrt{0.3}~i$,
$\sin\theta=\sqrt{1.3}$ and $(x_{11}-x_1)/x_{11}=1/1.3$.
Using these parameters we can find that $|{\tilde U}_{\mu s}|^2\approx 0.0087$
and $|{\tilde U}_{es}|^2\approx 0.036$.
A prediction of this scenario is that $|{\tilde U}_{\tau s}|$ is
suppressed together with $|{\tilde U}_{\mu s}|$. 
Since 
$U_{\tau 1}U_{\tau 2}\approx -\sin^2\theta_{23}\sin\theta_{12}\cos\theta_{12}$
and
$U_{\mu 1}U_{\mu 2}\approx -\cos^2\theta_{23}\sin\theta_{12}\cos\theta_{12}$,
one can see in Eq. (\ref{Rmatrix4}) that
two terms contributing to $R_{\tau s}$ will cancel with each other
when two terms contributing to $R_{\mu s}$ cancel with each other.
For parameters shown above we have $|{\tilde U}_{\tau s}|^2\approx 0.0044$.

We see in the above example that realizing 
$|{\tilde U}_{es}|>|{\tilde U}_{\mu s}|$ in our model,
to be consistent with the evidences of sterile neutrino, leads to
a preference of IH than NH.
In more general case, one can find that this preference of IH
is also true. One can check that it's always difficult to
suppress the contributions proportional to $U_{e3}$ and $U_{\mu 3}$
in Eq. (\ref{Rmatrix3}) while making $|{\tilde U}_{e s}|$
and $|{\tilde U}_{\mu s}|$ of the order of magnitude of interests.
For IH there is no such a problem.

\vspace{0.3 cm}
 \noindent {\bf Conclusions}

In summary we have shown that seesaw mechanism plus FL symmetry 
provide a natural mechanism for having a light sterile neutrino.
A FL symmetry in right-handed neutrino sector requires that one 
of the three right-handed neutrinos is massless and decoupled
from all other neutrinos. With soft-breaking of FL symmetry in Majorana
mass sector, an eV scale right-handed neutrino coupled to
other light neutrinos can emerge
and it can play the role of eV scale sterile neutrino required
for explaining experiments such as LSND, MiniBooNE, reactor flux
anomaly and Gallium radioactive source experiment. 
We solve the Yukawa coupling terms for the case with soft-breaking
of FL symmetry and find that the mass squared differences and mixing 
angles of active neutrinos can be easily accommodated in this framework.

We find that one light neutrino has to be massless and
the mass pattern of active neutrinos is either NH or IH.
Mixing of active neutrinos with sterile neutrino can be computed
using the Yukawa couplings solved for explaining the mass squared
differences and mixings of active neutrinos. Interestingly,
we find that the evidences of sterile neutrino prefer to have IH 
of active neutrinos in our model. We find that
for NH it is difficult have $|{\tilde U}_{e s}|>|{\tilde U}_{\mu s}|$ 
which is preferred by the evidences of sterile neutrino.
For IH we have shown it is not hard to accommodate this hierarchy
in our model of sterile neutrino.

We give an explicit example which gives 
a nice explanation of the
best fit of the sterile neutrino mass and the mixing with active neutrinos.
We find that realizing relatively large $|{\tilde U}_{es}|$
and relatively small $|{\tilde U}_{\mu s}|$ in our model naturally
leads to relatively small $|{\tilde U}_{\tau s}|$.
This interesting prediction can be tested in future 
atmospheric or solar neutrino experiments.

\acknowledgments
This work is supported by National Science Foundation of
 China(NSFC), grant No.11135009, No. 11375065 and Shanghai Key Laboratory
 of Particle Physics and Cosmology, grant No. 11DZ2230700.


\end{document}